\input{aipcheck}

\documentclass[
    ,final            
  ]
  {aipproc}

\layoutstyle{8x11single}
\usepackage{multirow}

\begin{document}

\title{Structure of the lightest scalar meson from the 1/$N_c$ expansion of
  Unitarized Chiral Perturbation Theory and Regge Theory}


\author{J. Ruiz de Elvira}{
  address={Departamento de F\'isica Te\'orica, II (M\'etodos Matem\'aticos), Facultad de Ciencias F\'isicas, Universidad
Complutense de Madrid, E-28040, Madrid, Spain}
}

\author{J.R. Pelaez }{
  address={Departamento de F\'isica Te\'orica, II (M\'etodos Matem\'aticos), Facultad de Ciencias F\'isicas, Universidad
Complutense de Madrid, E-28040, Madrid, Spain}
}

\author{M.R. Pennington}{
  address={Institute for Particle Physics Phenomenology,
Physics Department, Durham University, Durham DH1 3LE, U.K.}
}

\author{D.J. Wilson}{
  address={Institute for Particle Physics Phenomenology,
Physics Department, Durham University, Durham DH1 3LE, U.K.} 
}

\begin{abstract}
One-loop unitarized Chiral Perturbation Theory (UChPT) calculations, suggest a different Nc behaviour for
the $\sigma$ or $f_0$(600) and $\rho$(770) mesons \cite{Pelaez:2003dy}: while the $\rho$ meson becomes
narrower with Nc, as is expected for a $\bar{q}q$ meson, the $\sigma$ becomes
broader, and its contribution to the total cross section is less and less
important.  
On the other hand, local duality requires a cancellation between the $\sigma$
and $\rho$ amplitudes, but if there is a different Nc behaviour for
them, there is a possible contradiction between the Inverse Amplitude Method (IAM) and local duality for
large Nc.
However, next to next to leading order UChPT calculations suggested a subdominant
$\bar{q}q$ component for the $\sigma$ with a mass around 1.2 GeV \cite{Pelaez:2006nj}.
In this work, we show that this subdominant $\bar{q}q$ component is indeed needed to ensure
local duality. 
\end{abstract}

\maketitle


Light hadron spectroscopy lies outside the applicability range of QCD
perturbative  calculations. Still, in this low energy region one can use Chiral
Perturbation Theory (ChPT) \cite{Weinberg:1978kz} to obtain a model independent description
of the dynamics of pions, kaons and etas. These particles are the Goldstone
Bosons (GB) associated to the QCD spontaneous breaking of Chiral Symmetry
and ChPT is built as a low energy expansion that contains those fields
in the terms of a Lagrangian that respect all QCD symmetries, including its
symmetry breaking pattern. The small quark masses of the three lightest flavours
can be treated systematically within the perturbative chiral expansion
and thus ChPT becomes a series in momenta and meson masses, generically
$O(p^2/\Lambda^2)$. At lowest order there are no free parameters apart from masses
and $f_\pi$, the pion decay constant, that sets the scale $\Lambda \equiv
4\pi f_\pi$. 
The chiral expansion can be renormalized order by order by absorbing the loop divergences
in higher order counterterms, known as low energy constants (LEC),
whose values depend on the specific QCD dynamics. 

The renormalized LEC have to be determined from experiment, since they
cannot be calculated from perturbative QCD. However, thanks to the fact
that ChPT has the same symmetries than QCD and that it should couple
to different kind of currents in the same way, it is still possible to determine
in a model independent way how the constants that appear in ChPT, and
therefore the observables, depend on some QCD parameters. This is indeed
the case of the leading dependence on the number of colors Nc.

\subsection{Unitarization and dispersion theory}
The unitarity of the S matrix implies that, for physical values of s, partial
waves $t^{IJ}$ of definite isospin I and angular momenta J for \textit{elastic} meson-meson
scattering should satisfy:
\begin{equation}\label{unitarization}
  \rm{Im}\;t^{IJ}=\sigma
  |t^{IJ}|^2\;\;\;\Rightarrow\;\;\;Im\frac{1}{t^{IJ}}=-\sigma\;\;\;\Rightarrow\;\;\;t=\frac{1}{\mathrm{Re}\,t^{-1}-\mathrm{i}\sigma}
\end{equation}
where $\sigma = 2p/\sqrt{s}$, and p is the CM momenta of the two mesons. Note that
unitarity implies that $|t^{IJ} | \leq 1/\sigma$, and a strong interaction is characterized
precisely by the saturation of this unitarity bound.

However, partial waves are obtained within ChPT as a low energy expansion
$t \simeq t_2+t_4+t_6+\cdots$, (To simplify the notation, from now on we will drop
the IJ indices.) where $t_{2k} \equiv O(p/(4\pi f_\pi))^{2k}$, and thus they cannot satisfy
unitarity exactly, but just perturbatively, i.e: $\mathrm{Im}\,t_2 = 0,\;\;\; \mathrm{Im}\,t_4 = \sigma t_2^2,\;\;\;\; etc\, ...$

Unitarization methods provide amplitudes up to higher energies by using
the fact implicit in Eq. \eqref{unitarization}, that \textit{the imaginary part of the inverse amplitude
is known exactly}. We can then impose the ChPT constraints through the real part of $\rm{Re}\,t^{-1}
\simeq t^{-2}_2 (t_2 + \rm{Re}\,t_4 + ...)$ to find that
\begin{equation}
t = \frac{1}{\rm{Re}t^{-1}- i\sigma}\simeq\frac{t_2}{1-t_4/t_2}.
\end{equation}

This is known as the one-channel Inverse Amplitude Method (IAM) \cite{Truong:1988zp,Dobado:1992ha}. A usual
complaint is that the ChPT series is only valid at low energies, and there is
no reason to use it beyond that regime. However note that this formula can be 
derived from a dispersion relation \cite{Dobado:1992ha}, with an exact elastic cut, 
where ChPT is only used for the subtraction constants and the left cut in the low
energy region.  The simple formula of the IAM, Eq. (3), is able to describe
the meson-meson scattering data not only at low energies, where it reproduces
the ChPT series, but also in the resonance region. This is done with values
of the ChPT parameters that are compatible with the values obtained within
standard ChPT.
In addition, the IAM generates the poles [3,4] associated with the resonances
in the second Riemann sheet. This is of relevance since, in particular, the
scalar resonances are the subject of intense debate that has been lasting for
several decades, and as we have seen, the IAM is able to generate their poles
from first principles like unitarity, analyticity and the QCD chiral symmetry
breaking, without introducing these resonances by hand. Thus, we can study,
without any \textit{a priory} assumption, the nature of these states as follows from
first principles and QCD.

\subsection{The 1/Nc expansion}
The QCD 1/Nc expansion \cite{'tHooft:1973jz} allows for a clear identification of a $\bar{q}q$ resonance,
since it becomes a bound state, whose width follows an O(1/Nc)
behaviour, whereas its mass should behave as O(1). For our purposes, the relevant
observation is that the leading 1/Nc behaviour of the ChPT constants is
known in a model independent way. Thus, in order to know the leading Nc
behaviour of the resonances generated with the IAM, we just have to change
the ChPT parameters according to their established Nc scaling properties.

This procedure \cite{Pelaez:2003dy} was first applied to the one-loop SU(3) ChPT amplitudes, 
and the result was that the light vector resonances, as for example the $\rho$(770) followed the
expected behaviour of $\bar{q}q$  states remarkably well.
In contrast, the members of the light scalar nonet, like the $\sigma$ or
$f_0$(600), showed a behaviour at odds with that of $\bar{q}q$ states, Figure 1.
It follows that the dominant component of light scalar mesons does not have a $\bar{q}q$ nature.

   \begin{figure}[t]
     \centering
     \includegraphics[height=.25\textheight,angle=-90]{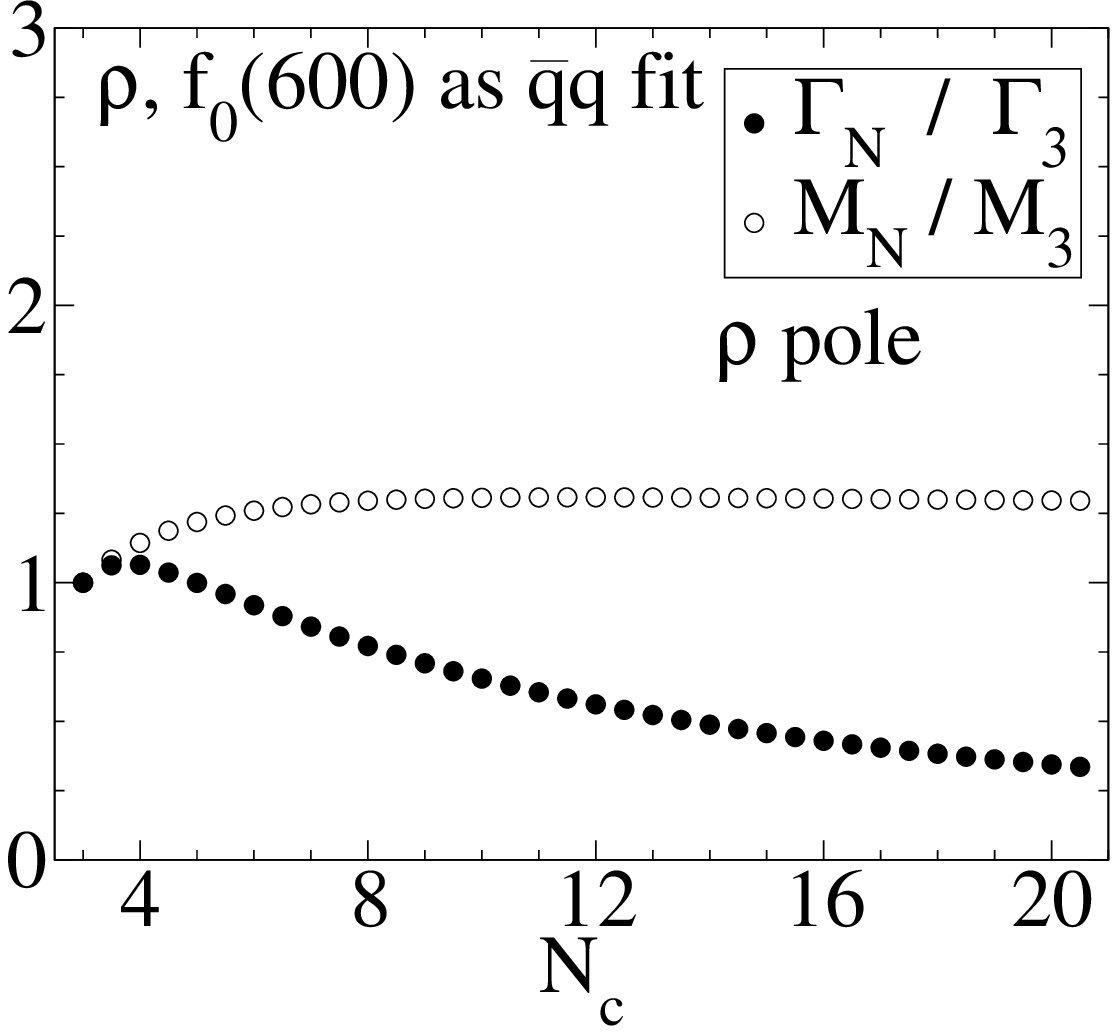}
     \includegraphics[height=.27\textheight,angle=-90]{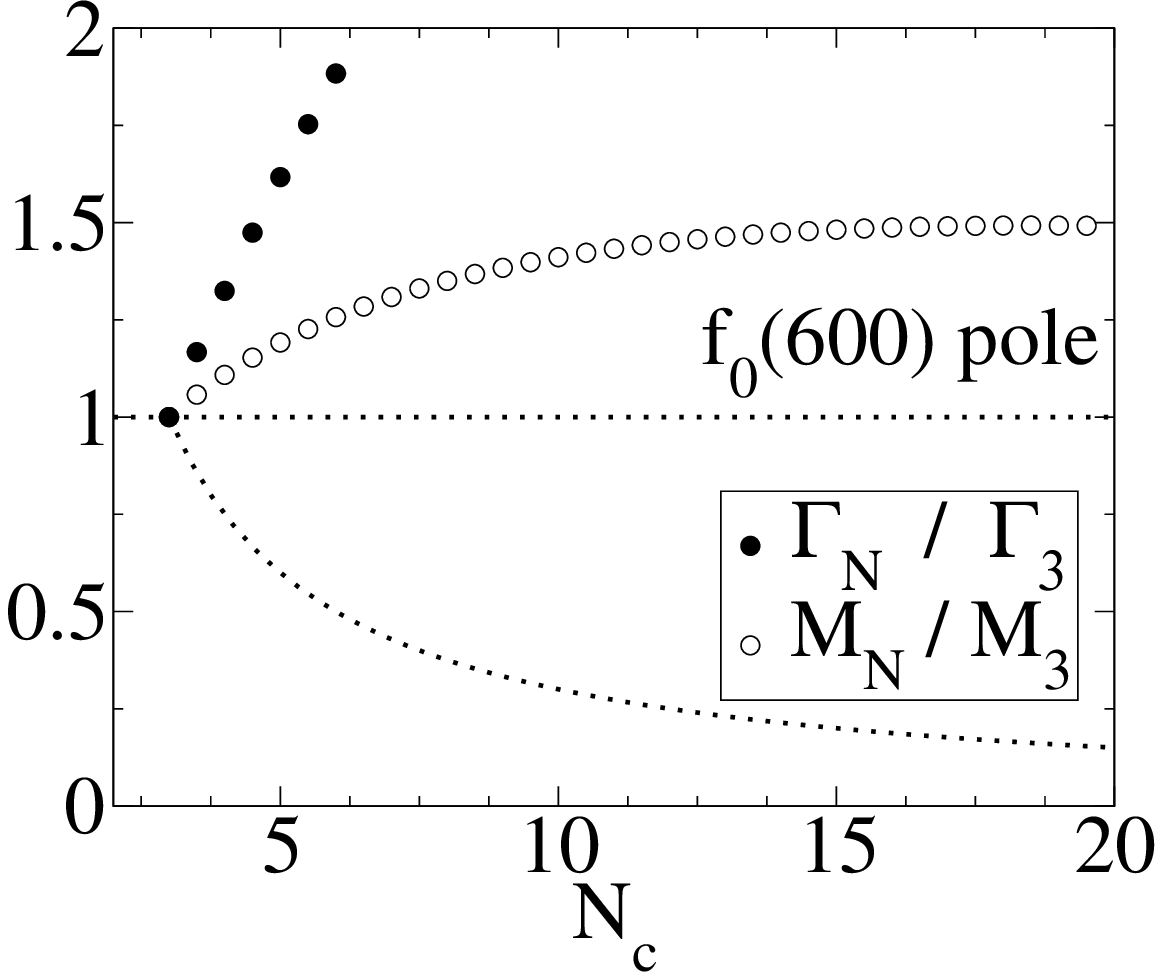}
     \includegraphics[height=.25\textheight,angle=-90]{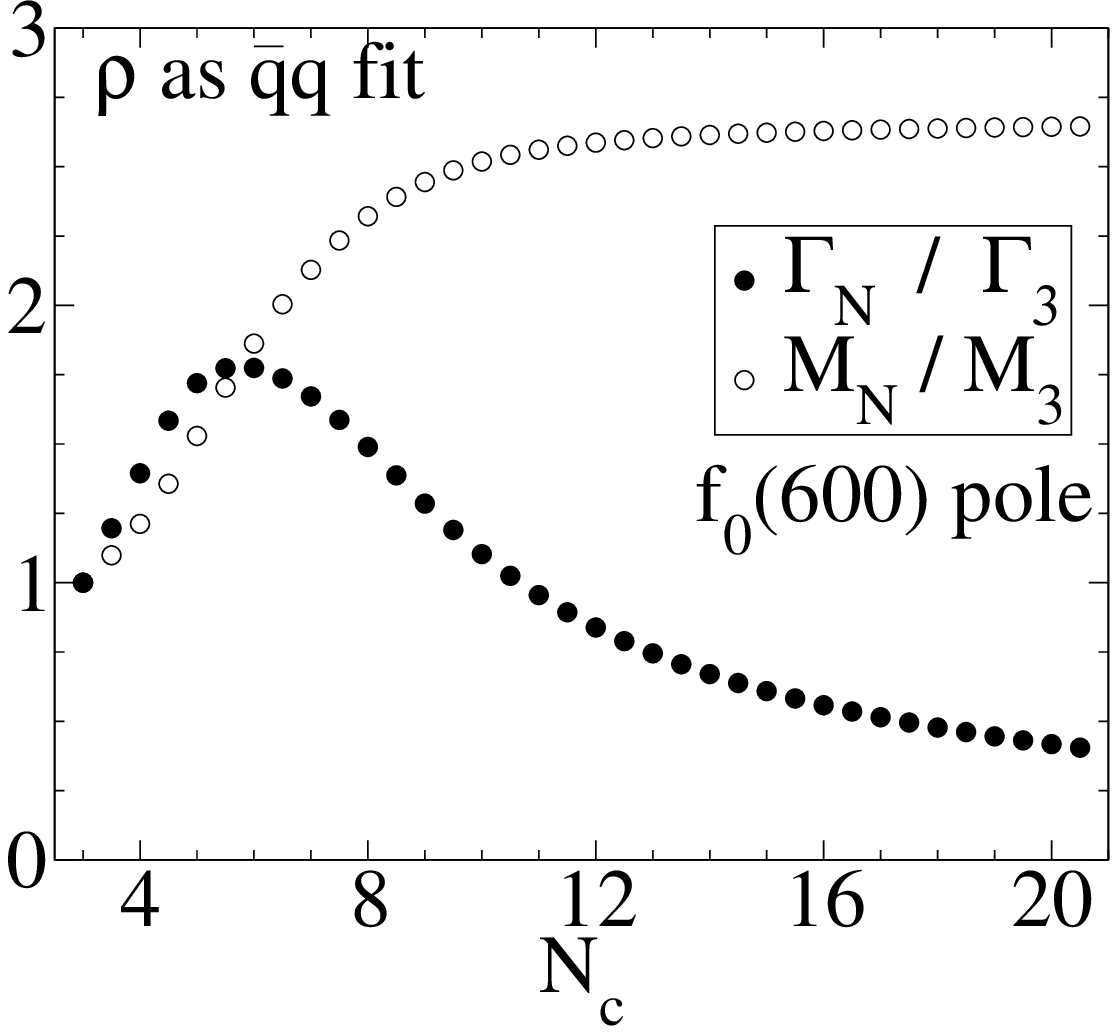} 
     \caption{Left, the $\rho$ meson behaves as a $\bar{q}q$ state. Center: At O($p^4$)
   the  $\sigma$ does not. Right: At O(p$^6$) emerges a subleading $\bar{q}q$
   component at 1.2 GeV}
 \label{fig:poles}
   \end{figure}

In \cite{Pelaez:2006nj} the two-loop SU(2) ChPT amplitude analysis 
has been performed; the results
show that the $\rho$(770) still follows naturally the expected $\bar{q}q$
behaviour. However if the $\rho$(770) is to behave as a
$\bar{q}q$, the $\sigma$ shows a different behaviour; namely, 
not far from Nc=3, both the $\sigma$ mass and width increase with the number of
colours, but from Nc $\sim$ 8 (in the best case), 
the mass of the $\sigma$ becomes constant and the width starts
decreasing with $\simeq$1/Nc, see Fig. 1. Right. 
In conclusion, the two-loop IAM
 confirms once again that the $\sigma$ or $f_0$(600) does not
behave predominantly as a $\bar{q}q$ state, but suggests the existence of a
subdominant $\bar{q}q$ component
that originates at a mass of $\simeq$ 1.2 GeV, which is approximately
twice that of the physical  $\sigma$ at $Nc=3$. This seems to support
models like \cite{vanBeveren:1986ea} that have indeed suggested 
a non-$\bar qq$ nonet below 1 GeV and an additional $\bar qq$ one
above.

\section{Local duality}
A long recognised feature of the world with Nc = 3 is that of ``local duality''. In a scattering
process, as the energy increases from threshold, distinct resonant structures give way to a smooth
Regge behaviour. At low energy the scattering amplitude is well represented by a sum of resonances
(with a background), but as the energy increases the resonances (having more phase space for decay)
become wider and increasingly overlap. This overlap generates a smooth behaviour of the cross-section
most readily described not by a sum of a large number of resonances in the direct channel, but the
contribution of a small number of crossed channel Regge exchanges. Indeed, detailed studies of meson-baryon
scattering show that the sum of resonance contributions at all energies ``averages'' (in
a well-defined sense to be specified below) the higher energy Regge behaviour. Thus, resonances in the s-channel
are related to Regge exchanges in the t-channel. Indeed, these resonance and Regge components
are not to be added like Feynman diagram contributions, but are ``dual'' to each other: one uses one
or the other. 

Regge exchanges are also built from $\bar{q}q$ and multiquark
contributions. In a channel like that with isospin 2 in $\pi\pi$ scattering there are no $\bar{q}q$  resonances, 
and so the Regge exchanges with these quantum number must involve multi-quark components. Data teach us that even
at Nc = 3 these components are suppressed compared to the dominant $\bar{q}q$ exchanges. Semi-local
duality means that in $\pi^{+}\pi^{-} \rightarrow \pi^{-}\pi^{+}$ scattering, the low energy resonances must have contributions to
the cross-section that ``on the average'' cancel, since this process is purely isospin 2 in the t-channel.

Using the crossing relationships we can express the I=2 t-channel amplitude as
a function of s-channel amplitudes: 
\begin{equation}
\mathrm{Im}\,A^{t2}(s,t)=\frac{1}{3}\mathrm{Im}\,A^{s0}(s,t)-\frac{1}{2}\mathrm{Im}\,A^{s1}(s,t)+\frac{1}{6}\mathrm{Im}\,A^{s2}(s,t),
\end{equation} since the $A^{s2}$ amplitude is repulsive and its contribution
is small, this suppression comes from a strong cancellation between the
$A^{s0}$ and $A^{s1}$.

Now in $\pi\pi$ scattering below 900 MeV, there are for I=0,1 just two low energy
resonances: the $\rho$(770) I=J=1 and the $\sigma$ or $f_0$(600) I=J=0.
Local duality requires that the contribution of these two components to the $\pi^{+}\pi^{-}$ cross section
do indeed ``on average'' cancel in keeping with I = 2 exchange in the
t-channel. 

This ``on the average cancellation'' is properly defined via Finite Energy Sum
Rules:   
\begin{equation}\label{FESR}
F(t)_n= \frac{\int_{\nu_{th}}^{\nu_{\mathrm{max}}}{d\nu\;\mathrm{Im}\, A^{t2}(s,t)/\nu^n}}{\int_{\nu_{th}}^{\nu_{\mathrm{max}}}{d\nu\;\mathrm{Im}\,A^{t1}(s,t)/\nu^n}},\;\;\;\nu=(s-u)/2.
\end{equation}

Semi-local duality between Regge and resonance contributions teaches us that
on the ``average'' and at least over one resonance tower, we have:
\begin{equation}\label{ReggeLocal}
  \int_{\nu_{\mathrm{th}}}^{\nu_{\mathrm{max}}}{d\nu\;\nu^{-n}\mathrm{Im}\,A^{t2}(s,t)_{\mathrm{Data}}}\sim \int_{\nu_{\mathrm{th}}}^{\nu_{\mathrm{max}}}{d\nu\;\nu^{-n}\mathrm{Im}\,A^{t2}(s,t)_{\mathrm{Regge}}},
\end{equation}
where Regge amplitudes are given by:\begin{equation}\label{Regge}
\mathrm{Im}\,A^{tI}=\sum_{R}
\beta_{R}(t)[\alpha^{\prime}(\nu-\nu_{th})]^{\alpha_R(t)}, 
\end{equation}where $\alpha_R(t)$ denotes the Regge trajectories with the appropriate t-channel quantum numbers, $\beta_R(t)$
their Regge couplings and $\alpha^{\prime}$ is the universal slope of the $\bar{q}q$ meson trajectories ($\sim$ 0.9 GeV$^{-2}$). $\nu_{th}$ is the
value of $\nu$ at threshold. 
Thus for the amplitude with I = 1 in the t-channel, the sum will be dominated
by $\rho$-exchange with a trajectory $\alpha(t) = \alpha_0 + \alpha^{′}$ t
that has the value 1 at $t = m_\rho^2$ and 3 at $t = m_{\rho 3}^2$ ,
i.e. $\alpha_0 \sim 0.467$ and  $\alpha_0^\prime \sim 0.889$ GeV$^{−2}$. 
 For the exotic I = 2 channel, we expect $\alpha(0) \ll \alpha_\rho(0)$, and its couplings to be
correspondingly smaller. Therefore using \eqref{ReggeLocal} and \eqref{Regge}, local duality implies
that $|F(t)_n| \ll 1$


As we have said above, using the IAM it is possible to check the dependence on Nc of $\pi\pi$
scattering amplitudes. In particular the I=2 s-channel amplitude remains
repulsive with Nc, and still there is no resonance exchange. Therefore local
duality implies that the $t$-channel $I=2$ Regge exchange should continue to
be suppressed as Nc increases. On the other hand the Regge trajectories do not depend on Nc, and therefore
$|F(t)_n| \ll 1$ when increasing Nc. As we have seen, this suppression comes from a strong cancellation between the
$\rho$(770) and the $\sigma$ or $f_0$(600) amplitudes.  The former with a
width of 170 MeV is a well established $\bar{q}q$  state, while the $\sigma$ is very broad, with a shape that is not Breit-Wigner-like,
and is not predominantly a $\bar{q}q$  state.
Such a distinct nature for the $\rho$(770) and $\sigma$ could prove a
difficulty as we increase Nc. This cancellation has to be checked at different number of colors.

\section{Results}

\begin{table}
    \begin{tabular}{c|c||c|c|c}
      $\nu_{max}$ & 400 GeV$^2$ &2.5 GeV$^2$&2 GeV$^2$&1 GeV$^2$ \\\hline 
      $F_1$ &0.021 $\pm$ 0.016 & 0.180 $\pm$ 0.066 &0.199 $\pm$ 0.089 & -0.320 $\pm$ 0.007\\\hline
      $F_2$ &0.057 $\pm$ 0.024 & 0.068 $\pm$ 0.024 &0.063 $\pm$ 0.025 & -0.115 $\pm$ 0.013\\\hline
      $F_3$ &0.249 $\pm$ 0.021 & 0.257 $\pm$ 0.022 &0.259 $\pm$ 0.022 & 0.221  $\pm$ 0.021\\\hline
    \end{tabular}\label{table 1}
    \caption{Values of
        the ratio F$_n$ using the KPY parametrization and different cutoffs.
      All $F_n$ ratios for a 20 GeV cutoff turn out very small, of the order 1:50 or 1:15 . 
      However, we see that they are only 1:4 or 1:5 when $s_{max}$ is still
      $\sim$2 GeV$^2$}
\end{table}

Using the IAM we can study the behaviour of Eq. \eqref{FESR} with Nc, and
then check if $|F(t)_n| \ll 1$ when increasing the number of colors. 
However, the IAM is only valid in the low energy region,and we have to
check the influence of the high energy part on this cancellation.

For this reason we use the $\pi\pi$ KPY \cite{Kaminski:2006qe} data parametrization 
 as input and check the value of the FESR at
different \textit{cutoffs}.
In Table \ref{table 1} we check that local duality is satisfied for Nc=3 since $|F(t)_n|
\ll 1$. Besides for n=2,3 the main contribution to the FESR
suppression occurs below 1 GeV, the energy region where we can trust the chiral
expansion.
Therefore, we can use the IAM to study local duality, and check the FESR
suppression with Nc. 
In evaluating the amplitudes $\mathrm{Im}A^{sI}$, we represent these by a sum of s-channel
partial waves, so that:
\begin{equation}
  \mathrm{Im}\,A^{sI}(s,t)=\sum_{J}{(2J+1)\mathrm{Im}\,t^{IJ}(s)P_J(\cos(\theta_s))}.
\end{equation}
However, using the IAM only S0, P and S2 waves can be described. It it is
necessary to check the influence of those waves in Eq. \eqref{FESR}. In Table
\ref{table 2} we show how the influence of higher waves is around
10$\%$, and that the IAM predicts correctly the FESR suppression. 
\begin{table}
    \begin{tabular}{c|c|c|c}\hline
      \multicolumn{4}{c}{$\nu_{max}$=1 GeV$^2$} \\ \hline
      & KPY08& 1 loop UChPT  & 2 loop UChPT \\\hline\hline 
      F$_1$ & -0.350 $\pm$ 0.083&-0.322 $\pm$ 0.061& -0.351\\\hline
      F$_2$ & -0.131 $\pm$ 0.042&-0.122 $\pm$ 0.097& -0.172\\\hline
      F$_3$ & 0.215 $\pm$ 0.027 &0.206 $\pm$ 0.138 &0.145\\\hline
    \end{tabular}\label{table 2}
    \caption{ Comparison between the $F_n$, using only S and P waves with a \textit{cutoff} of 1 GeV$^2$, calculated with data
      parametrizations or our IAM amplitudes.}
\end{table}

Up to one loop, we have seen that the $\rho$(770) becomes narrower with Nc, as
expected of a $\bar{q}q$ resonance, but the $\sigma$ or $f_0$(600) becomes
broader, and its contribution to the total cross section is less and
less important with Nc. Therefore the $\rho$ will dominate the FESR when
increasing Nc, and the ratios $|F(t)_n| \rightarrow 1$.  Up to one loop there is a
conflict between the IAM and local duality. The $\sigma$ amplitude could not
completely vanish in order to fulfill this.

\begin{figure}
  \includegraphics[height=.17\textheight]{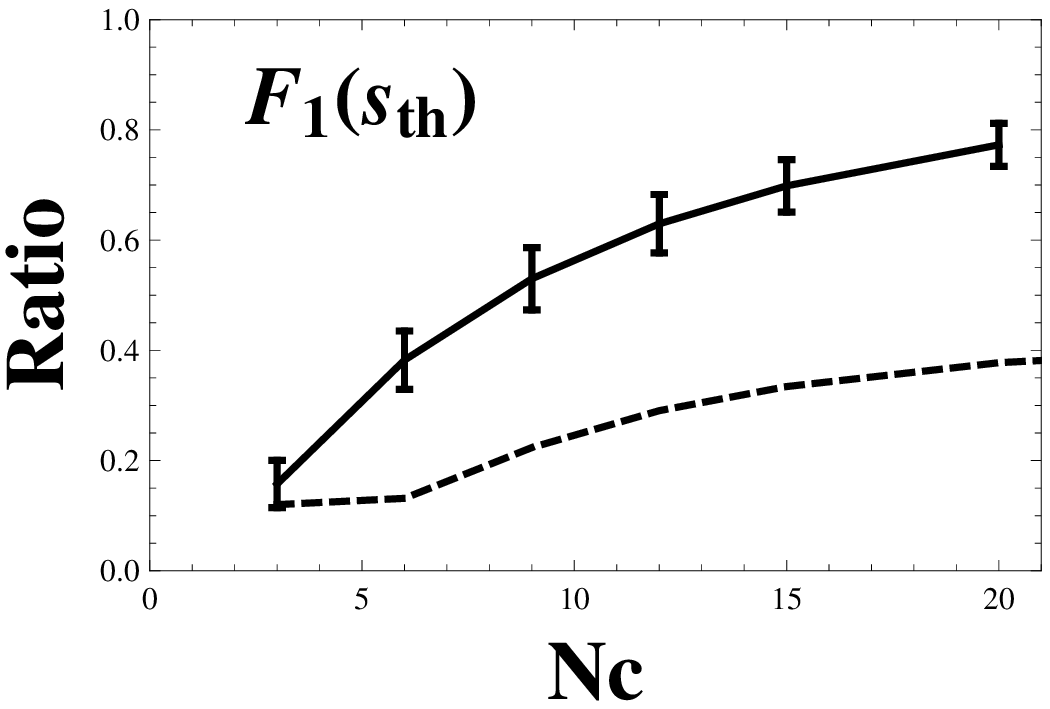}
  \includegraphics[height=.17\textheight]{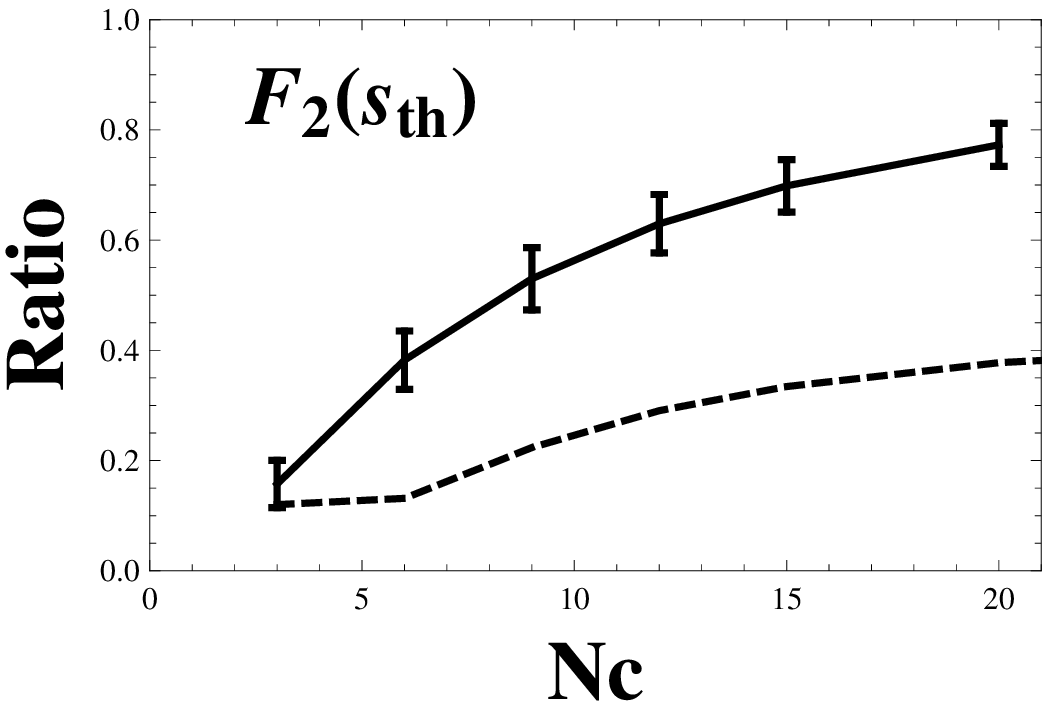}
  \includegraphics[height=.17\textheight]{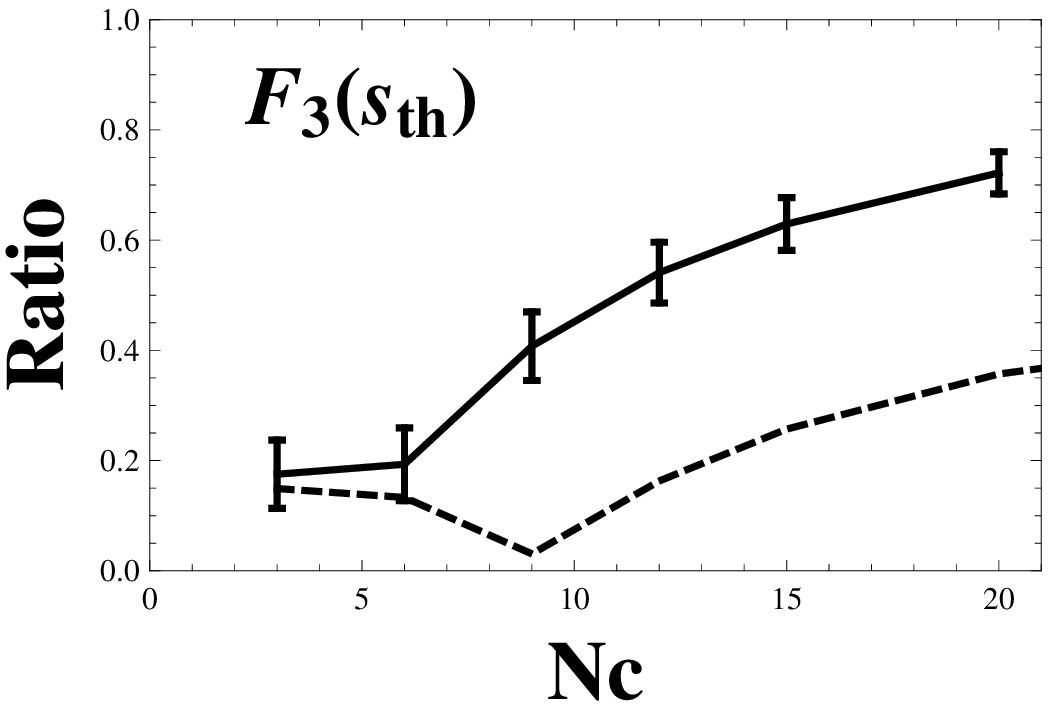}
  \caption{At O($p^4$), solid line, there is no FESR suppression and local duality
    fails as Nc grows. However, at O($p^6$), dashed line, 
the $\sigma$ subleading $\bar{q}q$ component ensures local
    duality even when increasing Nc}\label{Figure 3}
\end{figure}

However two-loop calculation seems to reveal a subdominant $\bar{q}q$
component for the $\sigma$ or $f_0$(600) with a mass around 1.2 GeV. 
Increasing Nc, there is still a cancellation between the $\sigma$ and the
$\rho$(770) amplitudes. In Table \ref{table 3}, we show both results. At one
loop, the ratios tend to one, but at two loops, they no longer increase so
much: the subleading $\bar{q}q$ component at 1 GeV ensures local duality.
In Figure \ref{Figure 3}, we show the difference between
|F($t_{th}$)$_1$|, |F($t_{th}$)$_2$| and |F($t_{th}$)$_1$| obtained using one-loop and two-loop UChPT.

\begin{center}
  \begin{table}
    \begin{tabular}{c|c|c}\hline
      n&Nc&$F(s_{\rm{th}})$ 1 loop\\\hline
      \multirow{4}{0.05in}{1} 
        & 3 & -0.355 $\pm$ 0.021\\
       & 6 & -0.438 $\pm$ 0.047\\
       & 9 & -0.538 $\pm$ 0.054\\
       & 12 & -0.621 $\pm$ 0.060\\\hline
      \multirow{4}{0.05in}{2} 
        & 3 &-0.157 $\pm$ 0.043\\
       & 6 &-0.382 $\pm$ 0.053\\
       & 9 &-0.530 $\pm$ 0.056\\
       & 12 &-0.630 $\pm$ 0.053\\\hline
      \multirow{4}{0.05in}{3}
        & 3 & 0.175 $\pm$ 0.062\\
       & 6 & -0.193 $\pm$ 0.066\\
       & 9 & -0.407 $\pm$ 0.062\\
       & 12 & -0.541 $\pm$ 0.055\\
    \end{tabular}
    \begin{tabular}{c}\hline 
         $F(s_{\rm{th}})$ 2 loops\\\hline
        -0.168\\
        -0.194\\
        -0.236\\
        -0.264\\\hline
        -0.120\\
        -0.131\\
        -0.223\\
        -0.290\\\hline
         0.149\\
         0.135\\
         -0.030\\
         -0.163\\
      \end{tabular}\label{table 3}
 \caption{ Two-loop IAM. The ratios grow only very slightly
when increasing Nc}
  \end{table}
\end{center}

\section{Conclusions}
The 1/Nc expansion of ChPT unitarized using the IAM shows that the $\sigma$ or
$f_0$(600) meson is not predominantly a $\bar{q}q$ state. Up to one loop, the
$\sigma$ amplitude vanishes, and there is a potential conflict between local duality and
the IAM at O($p^4$). However, the two-loop calculation reveals a subdominant
$\bar{q}q$ state that emerges around 1.2 GeV. This component ensures that local
duality is still satisfied as Nc increases within UChPT.

\section*{Acknowledgments}

Work partially supported by Spanish Ministerio de
Educaci\'on y Ciencia research contracts: FPA2007-29115-E,
FPA2008-00592 and FIS2006-03438,
U. Complutense/Banco Santander grant PR34/07-15875-BSCH and
UCM-BSCH GR58/08 910309. We acknowledge the support
of the European Community-Research Infrastructure
Integrating Activity
``Study of Strongly Interacting Matter''
(acronym HadronPhysics2, Grant Agreement
n. 227431)
under the Seventh Framework Programme of EU.

\end{document}